\documentclass[12pt]{article}
\usepackage{amsmath}
\usepackage{graphicx}
\usepackage{enumerate}

\usepackage{natbib,setspace,lscape,longtable,epsfig}
\usepackage{amsthm}
\usepackage{longtable}
\usepackage{array}
\usepackage{arydshln}

\usepackage{amsfonts}
\usepackage{mathrsfs}
\usepackage{multirow}
\usepackage{bbding}
\usepackage{amssymb}
\usepackage{amsmath}
\usepackage{graphicx,color}
\usepackage{rotating}
\usepackage{bm}
\usepackage{multirow}
\usepackage{tabularx}
\usepackage{booktabs}
\usepackage{epstopdf}
\usepackage{subfigure}
\usepackage{url}

\newcommand{\blind}{0}

\numberwithin{equation}{section}
\renewcommand{\baselinestretch}{1.75}

\DeclareMathOperator*{\argmin}{arg\,min}

\newcommand{\bea}{\begin{eqnarray}}
\newcommand{\eea}{\end{eqnarray}}
\newcommand{\Bea}{\begin{eqnarray*}}
\newcommand{\Eea}{\end{eqnarray*}}
\newcommand{\ba}{\begin{array}}
\newcommand{\ea}{\end{array}}
\newcommand{\bt}{\begin{tabular}}
\newcommand{\et}{\end{tabular}}
\newcommand{\btb}{\begin{table}}
\newcommand{\etb}{\end{table}}
\newcommand{\bc}{\begin{center}}
\newcommand{\ec}{\end{center}}
\newcommand{\beq}{\begin{equation}}
\newcommand{\eeq}{\end{equation}}

\usepackage{float}

\newtheorem{theorem}{\sc Theorem}[section]

\newtheorem{coro}{\sc Corollary}[section]

\makeatletter

\newcommand{\Rmnum}[1]{\expandafter\@slowromancap\romannumeral #1@}
\makeatother

\begin{document}

\def\spacingset#1{\renewcommand{\baselinestretch}%
{#1}\small\normalsize} \spacingset{1}


\if0\blind
{
  \title{\bf A Type of Nonlinear Fr\'{e}chet Regressions}
  \author{Lu Lin and Ze Chen\thanks{The corresponding author. Email: chze96@sdu.edu.cn.  Lin's research was supported by National Natural Science Foundation (NNSF) of China (No.11971265) and the National Statistical Science Research Project (2022LD03). Chen's research was supported by the Postdoctoral Fellowship Program of CPSF (GZC20231478).}
  \\
  \small Zhongtai Securities Institute for Financial Studies, Shandong University, Jinan, China\\
  }
  \maketitle
} \fi

\if1\blind
{
  \bigskip
  \bigskip
  \bigskip
  \begin{center}
    {\LARGE\bf Nonlinear Fr\'{e}chet Regressions}
\end{center}
  \medskip
} \fi

\bigskip
\begin{abstract}
The existing Fr\'echet regression is actually defined within a linear framework, since the weight function in the Fr\'echet objective function is linearly defined, and the resulting Fr\'echet regression function is identified to be a linear model when the random object belongs to a Hilbert space. Even for nonparametric and semiparametric Fr\'echet regressions, which are usually nonlinear, the existing methods handle them by local linear (or local polynomial) technique, and the resulting Fr\'echet regressions are  (locally) linear as well. We in this paper introduce a type of nonlinear Fr\'echet regressions. Such a framework can be utilized to fit the essentially nonlinear models in a general metric space and uniquely identify the nonlinear structure in a Hilbert space. Particularly, its generalized linear form can return to the standard linear Fr\'echet regression through a special choice of the weight function. Moreover, the generalized linear form possesses methodological and computational simplicity because the Euclidean variable and the metric space element are completely separable. The favorable theoretical properties (e.g. the estimation consistency and presentation theorem) of the nonlinear Fr\'echet regressions are established systemically. The
comprehensive simulation studies and a human mortality data analysis demonstrate that the new strategy is significantly better than the competitors.
\end{abstract}

\noindent%
\vfill

\newpage
\spacingset{1.9} 
\section{Introduction}

\subsection{Related issues and our goals}
When data types are out of the Euclidean framework and lack a general algebraic structure, the methods of metric space and Hilbert space in probability and statistics offer alternative tools for analyzing and modeling these abstract contexts. These methodologies shift the starting
point away from the Euclidean sample space, and replace it with a metric space or a Hilbert space according to the features of the abstract and complex elements. The early work of metric space methods in probability and statistics can date back at least to the Fr\'echet mean in \cite{Frechet:1948}, which is defined for random elements in a metric
space as the minimizer of an expected loss defined within that space. Similarly, the Hilbert space methods  employ notions such as inner product, orthogonality, and projection within linear subspaces, as alternative approaches to statistical inference on abstract and complex elements (see \cite{Small:Mcleish:1994} for the systematic introduction).

In recent years, the increasing complexity and abstraction of data types driven by application demands, the advancements in measurement techniques and improvements in data collection and storage, have sparked significant interest in statistical methodologies for analyzing random objects in non-Euclidean spaces (see, e.g., \cite{Marron:Alonso:2014}, \cite{Wang:Marron:2007} and \cite{Faraway:2014}). As a prominent issue in non-Euclidean space, the Fr\'echet regression was introduced by \cite{Petersen:Muller:2019}, in which the response is a random object defined in a metric space. Specifically, the Fr\'echet regression function is defined as the minimizer of an expected weighted loss defined within a metric space. The variable selection and model averaging for  Fr\'echet regression were further proposed by \cite{Tuckeretal:2023} and \cite{Yanetal:2023}, respectively.
Moreover, the Fr\'echet regression was recently extended into semiparametric models by \cite{Ghosal:Meiring:2023} and \cite{Bhattacharjee:Muller:2023}.

The existing Fr\'echet regressions, however, are actually defined within a linear framework. This is because the weight function in the Fr\'echet regression objective function is linearly defined, and the resulting Fr\'echet regression function is only suitable for the linear model. Even for nonparametric and semiparametric Fr\'echet regressions, which are often nonlinear, current methods handle them by local linear (or local polynomial) methods or approximately linear techniques, the resulting Fr\'echet regressions are locally or approximately linear as well (see, e.g., \cite{Petersen:Muller:2019}, \cite{Schotz:2022}, \cite{Ghosal:Meiring:2023} and \cite{Bhattacharjee:Muller:2023}).

To the best of our knowledge,  the pursuit of nonlinear Fr\'echet regression remains an open question. Constructing the framework of nonlinear Fr\'echet regression may be highly nontrivial. The main difficulties are as follows. In addition to that the framework should possess the common statistical properties, it should exhibit adaptability in the following aspects:
\begin{enumerate}
\item[(1)] it can correctly and even explicitly identify the underlying nonlinear regression function when the underlying regression is definitely nonlinear;
    \item[(2)] it should encompass a wide range of nonlinear functions, and yet return to a standard linear framework when the underlying regression is indeed linear;
    \item[(3)] it is easily implementable and computationally efficient especially for the case when the estimation procedure for the unknown parameters in a nonlinear function cannot be separated from the metric space.
\end{enumerate}
The challenges mainly come from the feature that the
random object in a metric does not live in a linear
space, thereby rendering common mathematical operations such as addition and multiplication inapplicable to these objects. This
feature results in a major problem for applying parameter based
methods to define the nonlinear Fr\'echet regression and estimate the unknown parameters. These challenges will be successively investigated in the following sections.

Besides, like the common regression in Euclidean space, the Fr\'echet regression is originally defined by a conditional expected loss \citep{Petersen:Muller:2019}. The classic approach to tackling the conditional expectation problem  involves transforming it into a system of unconditional expectations. Then, the unconditional expectation methods, such as the optimally weighted generalized method of moments, can be employed to realize statistical inference.  This transformation is mathematically convenient, as the objective function constructed from unconditional moments can be analytically computed. Early work on this approach can be found in \cite{Hansen:1982}, and present variants are discussed in \cite{Bennett:Kallus:2023}, among many others.
Particularly, when the unconditional moments are defined in a product space of Euclidean and metric spaces, it is expected that the variables in Euclidean space and the objects in metric space can be separated into different factors such that the procedures of modeling and analyzing can be implemented separately in the respective spaces.

In this paper, our goal is to establish the framework of nonlinear Fr\'echet regression. The key is to elaborately design the related Fr\'echet objective functions in the following form: they are expected weighted losses, in which the weights are a type of nonlinear functions of covariates defined in an Euclidean space, and the loss is the distance defined in a metric space. Consequently, the  Fr\'echet regression functions can be defined as the minimizer of these Fr\'echet objective functions. Such a structural nonlinear framework is essentially different from that of linear Fr\'echet regression. The new proposals can be utilized to fit true nonlinear models, and their separable structures can be easily analyzed and computed. Furthermore, the underlying nonlinear models can be explicitly identified by the proposed Fr\'echet regressions under a Hilbert space.
Particularly, the generalized linear Fr\'echet regression can return to the standard linear Fr\'echet regression through a special choice of the weight functions. Theoretically, some favorable properties, such as the estimation consistency and the convergence rate, can be established systematically.

\subsection{Problem setup}
Let $\Omega$ be a metric space, equipped with a metric $d$. The
Fr\'{e}chet regression function of an object response $Y\in \Omega$
on a covariate $X \in R^p$ is defined as the minimizer of the conditional expectation of a loss defined in the metric space \citep{Petersen:Muller:2019}. Formally, the Fr\'{e}chet regression function is originally written as
\begin{equation}\label{(definition)}m_\oplus(x)=\argmin_{\omega\in \Omega}E[d^2(Y,\omega)|X=x].
\end{equation}
 However, it is difficult to solve the conditional expectation problem, especially for the issue in a metric space. Just like the classical strategies for the problems in Euclidean spaces,
an approach to the conditional expectation in a metric space is to reduce it to an unconditional expectation with some weight factors (see, e.g., \cite{Hansen:1982} and \cite{Bennett:Kallus:2023}. Then, in order to easily implement the Fr\'{e}chet regression, with $\mu=E[X]$ and $\Sigma=Var[X]$, the existing literature \citep{Petersen:Muller:2019} utilizes the weight
$s(X,x)=1+(X-\mu)^T\Sigma^{-1}(x-\mu)$
to reconstruct the Fr\'{e}chet regression function as the minimizer of the unconditionally expected weighted loss; namely, the Fr\'{e}chet regression function is equivalently rewritten as
\begin{equation}\label{(expression)}m_\oplus(x)=
\argmin_{\omega\in\Omega}E[s(X,x)d^2(Y,\omega)].\end{equation}

In the above, the weight function $s(X,x)$ is in a parameter framework and exhibits linearity in $x$.
When the Fr\'{e}chet regression is redefined by the above unconditional expectation, the resulting model is indeed linear in $x$ with the structure similar to the model (\ref{(linear-model)}) under a Hilbert space (see Theorem 5 of \cite{Petersen:Muller:2019}). For emphasizing the linearity and differentiating it from the nonlinearity studied in this paper, {\it here we call $m_\oplus(x)$ the linear Fr\'{e}chet regression function, denoted by LFR for short, and refer to $E[s(X,x)d^2(Y,\omega)]$ as the linear Fr\'{e}chet objective function}.
By the unconditional expectation expression, the LFR function can be computed analytically for the case where the Euclidean variable $X$ and the metric space object $Y$ are separated into different factors $s(X,x)$ and $d^2(Y,\omega)$.

The LFR function defined in (\ref{(expression)}) always passes through the central point $(\mu,\omega_\oplus)$, since $s(\cdot,\mu)\equiv1$ implies that $m_\oplus(\mu) = \omega_\oplus$, where $\omega_\oplus =\argmin_{\omega\in\Omega}E[d^2(Y,\omega)]$ is the Fr\'echet mean of $Y$ \citep{Frechet:1948}.
Particularly, consider the
common linear regression model in an Euclidean space defined by
\begin{equation}\label{(linear-model)}Y=\beta^{(0)}+\beta^T(X-\mu)+\varepsilon,
\end{equation}
where the response $Y\in R$, $\mu=E[X]\in R^p$, $\beta^{(0)}\in R$ and $\beta\in R^p$. 
Under the above linear model defined in an Euclidean space, the LFR function defined in (\ref{(definition)}) happens to be the minimizer of the unconditional weighted expectation given in (\ref{(expression)}) with the linear weight $s(X,x)$.

\begin{enumerate} \item[]
{\it However, it is intuitive that if the underlying model is actually nonlinear, the regression function should NOT always pass through the central point $(\mu,\omega_\oplus)$. A typical example is that if $m_\oplus(x)=m(x)$ is a strictly convex function, then it always holds that $\omega_\oplus>m(\mu)$, which is against the rule of passing through the central point $(\mu,\omega_\oplus)$. Furthermore, a linear Fr\'{e}chet objective function uniquely identifies a linear regression function in a Hilbert space. Thus, if the underlying model is nonlinear, the weight in the objective should NOT always be of the linear framework. 
}
\end{enumerate}

The above discussions show that under the nonlinear case, the framework of Fr\'echet regression should have essential differences from those in the linear case.
These observations motivate us to develop new methodology and theory for constructing nonlinear Fr\'echet regression.

\subsection{Article structure}

The remainder of the paper is then organized in the following way. In Section 2, the nonlinear models in Euclidean space are first reexamined to motivate the development of the nonlinear model in a metric space. Then, the definition of nonlinear Fr\'echet regression is proposed, and particularly, a generalized linear Fr\'echet regression is introduced. The related representation theorems are investigated. For illustration, some important examples are discussed. In Section 3, the estimation methods and the algorithms are presented, and the asymptotic properties are established. The main simulation studies and the main result of a human mortality data analysis are presented respectively in Section 4 and Section 5. Section 6 concludes the paper
and discusses the possible future works. The Supplementary Materials include the following materials: the regularity conditions and the proofs for lemmas and theorems, some implementation details, and some additional simulation results and the details of the real data analysis.

\section{Nonlinear Fr\'echet regression modeling}

Here, our goal is to establish the framework of nonlinear Fr\'echet regression and investigate its basis properties.

\subsection{A type of nonlinear Euclidean regressions}

It is well-known that nonlinear regression is widely used in various fields. A famous example of nonlinear regression is the Gompertiz model defined by
\begin{equation}\label{(Gompertiz)}
Y=\theta_1\exp{[-\exp(\theta_2-\theta_3^{X})]}+\varepsilon.
\end{equation}
This is a sigmoidal growth model and commonly appears in some application fields such as biology and economics. For more examples of nonlinear regression models, particularly in biostatistical contexts, see Ratkowsky (1983), Seber and Wild (2003) and Wei (1998), among others.
Generally, a nonlinear regression model has the following unified form:
\begin{equation}\label{(nonlinear-regression)}Y=f(X,\beta)+\varepsilon
\end{equation}
for a given function $f$ with covariate $X\in R^p$ and unknown parameter $\beta\in R^q$.
Particularly, when $p=q$, a special structure for the regression function is $f(x,\beta)=g(\beta^T(x-\mu))$, where $g$ is a known link function. In this case, the resulting model is a generalized linear model, that is,
\begin{equation}\label{(generalized-regression)}
Y=g(\beta^T(X-\mu))+\varepsilon. \end{equation}
Under an exponential family distribution, the link function $g$ has a canonical or natural choice.
It is known that generalized linear regression is one of the most extensive models in the field of statistics (see, e.g., \cite{McCulloch:Searle:2001} and \cite{Hardleetal:2004}).

Denote $X=(X^{(1)},\cdots,X^{(p)})^T$
and $\mu=(\mu^{(1)},\cdots,\mu^{(p)})^T=E[X]$, and suppose that $\Sigma=Var[X]$ exists and is positive.
To motivate the methodological development, we consider a type of nonlinear Euclidean regression of $Y\in R$ on $X\in R^p$ as
\begin{equation}\label{(nonlinear-function)}
m^{\cal N}(x,\beta)=E[Y|X=x]=E[Y]+\sum_{j=1}^{p}\sigma^{(j)}
f_j\left(x,\beta\right),
\end{equation} where $\beta\in \mathscr B$ is an unknown parameter vector, the parameter space $\mathscr B$ is a subset of $R^q$, $f_j$'s are known regression functions,
and $\sigma^{(j)}=E[Y(X^{(j)}-\mu^{(j)})]$. The above model (\ref{(nonlinear-function)}) is a special case of general regression (\ref{(nonlinear-regression)}), but it is a big class of nonlinear regressions duo to the arbitrariness of the choices of $f_j$, and it contains the famous Gompertiz model  (\ref{(Gompertiz)}) as its special case. Also the model can be derived from exponential families nonlinear models (see, e.g., \cite{Wei:1998}). The first condition for nonlinear modeling is the assumption of $\sigma^{(j)}\neq 0$ for some $j$. This condition is easy to satisfy; otherwise, it can be replaced by $\sigma^{(j)}_t=E[Y(t(X^{(j)})-E[t(X^{(j)})])]\neq 0$ for some function $t$ and some index $j$. The condition $\sigma^{(j)}_t\neq 0$ is always satisfied if there exist some (nonlinear) correlations between the response $Y\in R$ and covariates $X^{(j)}\in R$.
For the nonlinear Euclidean regression, the nonlinear weight function is then defined by
\begin{equation}\label{weight-nonlinear}s^{\cal N}(X,x,\beta)=1+\sum_{j=1}^{p}(X^{(j)}-\mu^{(j)})f_j
(x,\beta).  \end{equation} The weight depends only on the Euclidean variables and parameters, and the known functions $f_j$, which is free of $\sigma^{(j)}$ although the underlying regression depends on $\sigma^{(j)}$.
Particularly, under the generalized linear case, the Euclidean regression function is designed as  \begin{equation}\label{(generalized-regression-function)}
m^{\cal N}(x,\beta)=E[Y|X=x]=E[Y]+\sum_{j=1}^{p}\sigma^{(j)}
f_j\left(\beta^T(x-\mu)\right), \end{equation} and the corresponding generalized linear weight function is defined by
\begin{equation}\label{generalized-weight-nonlinear}s^{\cal N}(X,x,\beta)=1+\sum_{j=1}^{p}(X^{(j)}-\mu^{(j)})f_j
(\beta^T(x-\mu)).  \end{equation} As aforementioned, the model \eqref{(generalized-regression-function)} can be derived from the exponential families generalized linear models.

Since $E[s^{\cal N}(X,x,\beta)]=1$ for all $x$ and $\beta$, the regression function $m^{\cal N}(x,\beta)$ defined in (\ref{(nonlinear-function)}) and (\ref{(generalized-regression-function)}) can be recast as
\begin{equation}\label{weight-sum}\begin{split}
m^{\cal N}(x,\beta)=
E\left[s^{\cal N}(X,x,\beta)Y\right], \end{split}\end{equation} which is the weighted expectation of $Y$ with the weight function $s^{\cal N}(X,x,\beta)$ as in (\ref{weight-nonlinear}) or (\ref{generalized-weight-nonlinear}).
Then, under Euclidean space, the nonlinear regression function $m^{\cal N}(x,\beta)$ in (\ref{(nonlinear-function)}) or (\ref{(generalized-regression-function)}) is indeed the solution to the following optimization problem:
\begin{equation}\label{(nonlinear-expression)}m^{\cal N}(x,\beta)=
\argmin_{y\in R}E[s^{\cal N}(X,x,\beta)d_E^2(Y,y)] \end{equation} for given $\beta$, where $d_E$ is the standard Euclidean metric.

\subsection{Nonlinear Fr\'echet regression}


The above alternative formulation (\ref{(nonlinear-expression)}) of the
nonlinear Euclidean regression function provides the key to defining the nonlinear Fr\'echet
regression function on an arbitrary metric space $(\Omega, d)$. By simply replacing
the Euclidean metric $d_E$ with
a more general metric $d$ that is suitable for responses in metric space $\Omega$, the nonlinear Fr\'echet
regression function is formally defined as
\begin{equation}\label{(nonlinear-Frechet)}m_\oplus^{\cal N}(x,\beta)=
\argmin_{\omega\in \Omega}E[s^{\cal N}(X,x,\beta)d^2(Y,\omega)] \end{equation} for given $\beta$ with the nonlinear weight function $s^{\cal N}(X,x,\beta)$ defined by (\ref{weight-nonlinear}) or (\ref{generalized-weight-nonlinear}). We call $m_\oplus^{\cal N}(x,\beta)$ {\it the nonlinear Fr\'echet regression function, denoted by NLFR for brevity}, because the weight function $s^{\cal N}(X,x,\beta)$ is in a nonlinear framework. A further reason is that it will be proved later that the NLFR can uniquely identify an underlying nonlinear regression function in a Hilbert space.
The pivotal features of the above definition are as follows:

\begin{enumerate}  \item[(1)] The loss (or the distance) $d^2(Y,\omega)$ is defined on the given metric space;
\item[(2)] The nonlinearity of the regression function $m_\oplus^{\cal N}(x,\beta)$ is completely determined by the nonlinearity of the weight function $s^{\cal N}(X,x,\beta)$ that is only related to Euclidean  variables;
\item[(3)] The Fr\'echet objective function is defined by an unconditional expectation, and the Euclidean variable $X$ and the metric space object $Y$ are separated into different factors $s^{\cal N}(X,x,\beta)$ and $d^2(Y,\omega)$.\end{enumerate}
In addition, it is necessary to achieve the adaptiveness to the linear case as in (\ref{(expression)}). In other words, with some special choices of the functions $f_j(\cdot)$ in the generalized linear case (\ref{generalized-weight-nonlinear}), the resulting model can be reduced to an LFR. To this end, let $\beta=
\Sigma^{-1}\sigma$ with $\sigma=(\sigma^{(1)},\cdots,\sigma^{(p)})^T$, and rewrite $f_j$ as
$$f_j\left(\sigma^T
\Sigma^{-1}(x-\mu)\right)=f_j
\Big(\sum_{i=1}^{p}\sigma^{(i)}\sum_{k=1}^{p}v^{(ik)}(x^{(k)}-\mu^{(k)})\Big)$$
with $(v^{(ik)})_{i,k=1}^{p}=\Sigma^{-1}$.
If each link function $f_j(\cdot)$ is chosen as the following linear function:
\begin{equation}\label{fj}f_j(u)=\frac{u-\sum_{i\neq j,i=1}^{p}\sigma^{(i)}\sum_{k=1}^{p}v^{(ik)}(x^{(k)}-\mu^{(k)})}{\sigma^{(j)}},
 \end{equation} then, $f_j\left(\sigma^T
\Sigma^{-1}(x-\mu)\right)$ has the following linear representation:
$$f_j\left(\sigma^T
\Sigma^{-1}(x-\mu)\right)=\sum_{k=1}^{p}v^{(jk)}(x^{(k)}-\mu^{(k)}).  $$
We then have the following finding:
\begin{enumerate} \item[]{\it
Based on the choice of $f_j(\cdot)$ in (\ref{fj}), the generalized linear weight $s^{\cal N}(X,x,\beta)$ given in (\ref{generalized-weight-nonlinear}) reduces to the linear weight $s(X,x)$. Consequently, the resulting model returns to the standard LFR defined in \cite{Petersen:Muller:2019}.}\end{enumerate}

Next, we study the basic properties of the NLFR and the representation theorems. It is clear that the newly defined NLFR has the following favorable properties:
\begin{enumerate} \item[(1)]
When the NLFR reduces to an LFR (i.e., all $f_j,j=1,\cdots,p$, are given in (\ref{fj})), the regression curve always passes through the central point $(\mu,\omega_\oplus)$.
\item[(2)]
When the NLFR function is essentially nonlinear (i.e., at least one of $f_j(\cdot),j=1,\cdots,p$, is different from the function in (\ref{fj})), the regression curve does not passes through the central point $(\mu,\omega_\oplus)$ except for the case of $f_j(0)=0$ for $j=1,\cdots,p$.
\item[(3)] The nonlinearity of the Fr\'echet regression is completely determined by the nonlinearity of the Euclidean weight function $s^{\cal N}(X,x,\beta)$.
\end{enumerate}
The above properties (1) and (2) ensure that the NLFR is adaptive to ``the center rule" according to whether the Fr\'echet regression is linear or not. The property (3) shows how to check the nonlinearity.

Moreover, another issue is about how to explicitly identify the NLFR function $m_\oplus^{\cal N}(x,\beta)$ when the random object $Y$ belongs to a Hilbert space. The following representation theorem states the details.

\begin{theorem}\label{Theorem 2.1.}  Suppose that $\Omega$ is a Hilbert space equipped with an inner product $\langle\cdot,\cdot\rangle$ and the corresponding norm $\|\cdot\|^2_\Omega$. If $E\|Y\|^2_\Omega <\infty$, then, there
exist unique elements $\beta^{(0)}\in \Omega$ and $\sigma^{(j)}\in \Omega$ such that
$E\langle Y,\omega\rangle=\langle\beta^{(0)},\omega\rangle$ and $E\langle (X^{(j)}-\mu^{(j)})Y,\omega\rangle=\langle\sigma^{(j)},\omega\rangle$ for all $\omega\in \Omega$.
With these elements $\beta^{(0)}$ and $\sigma^{(j)}$, the NLFR function $m_\oplus^{\cal N}(x,\beta)$ defined in (\ref{(nonlinear-Frechet)}) has the following explicit representation:
\begin{equation}\label{(representation)}
m^{\cal N}_\oplus(x,\beta)=\beta^{(0)}+\sum_{j=1}^{p}\sigma^{(j)}f_j
\left(x,\beta^T\right)\mbox{ or } m^{\cal N}_\oplus(x,\beta)=\beta^{(0)}+\sum_{j=1}^{p}\sigma^{(j)}f_j
\left(\beta^T(x-\mu)\right)\end{equation} for given $\beta$ according the weight function given by (\ref{weight-nonlinear}) or (\ref{generalized-weight-nonlinear}).
\end{theorem}

The representations in the theorem provide a convenient way for theoretical analysis and practical application.
Besides, similar to the Euclidean case as in (\ref{weight-sum}), the NLFR function can be expressed as a weighted expectation of $Y$. The following corollary gives the details.

{\coro \label{Corollary 2.1.} Under the conditions of Theorem \ref{Theorem 2.1.}, in a Hilbert space and for given $\beta$, the NLFR function $m_\oplus^{\cal N}(x,\beta)$ defined in (\ref{(nonlinear-Frechet)}) has the following representation:
\begin{equation}\label{(weighted-representation)}
m^{\cal N}_\oplus(x,\beta)=E[s^{\cal N}(X,x,\beta)Y]. \end{equation}
}

The corollary provides a new perspective to the regression function: under a Hilbert space, the NLFR function $m^{\cal N}_\oplus(x,\beta)$ can be regarded as a weighted expectation of the response $Y$ with nonlinear weight $s^{\cal N}(X,x,\beta)$. Thus, it could be expected that we can develop various Fr\'echet regressions by weighted expectation with various weights. We will continue to study this interesting issue in the future.

Next, we suggest the following examples that often appear in statistics to illustrate the conditions of Theorem \ref{Theorem 2.1.}.

{\it Example 1.} Let $\Omega$ be the set of quantile functions $G^{-1}$ of the probability distributions $G$ on $R$ such
that $\int_R x^2 dG(x) < \infty$, equipped with the Wasserstein metric $d=d_W$. Under Wasserstein metric, the Wasserstein distance is given by
$d^2_W(G_1,G_2)=\int_0^{1}(G^{-1}_1(t)-G^{-1}_2(t))^2dt$ for two
quantile functions $G_1^{-1}$ and $G_2^{-1}$ in $\Omega$. With these conditions, $\Omega$ is a Hilbert space. In this case, the resulting Fr\'echet regression function has the representations as in (\ref{(representation)}) and (\ref{(weighted-representation)}).

{\it Example 2.} Suppose that $\Omega$ is the set of correlation matrices of a fixed dimension
$r$, equipped with the Frobenius metric $d_F$. In this case, $\Omega$ is a Hilbert space, and the resulting Fr\'echet regression function has the representations as in (\ref{(representation)}) and (\ref{(weighted-representation)}) as well.

{\it Example 3.} It is known that the tangent vector field is often used to formalize the
notion of the Riemannian manifold. Let $\Omega$ be the tangent space of a bounded Riemannian manifold of dimension $r$, and
$d $ be defined by the inner product in the tangent space. In the Hilbert space $\Omega$, the resulting Fr\'echet regression function has the representations as in (\ref{(representation)}) and (\ref{(weighted-representation)}) as well.

These examples and the corresponding representations for the NLFR functions provide a basis for the simulation studies and real data analysis given in Sections 4 and 5.

\subsection{A separable representation for generalized linear form}

In the previous subsection, the definition and the representation theorem for the NLFR $m^{\cal N}_\oplus(x,\beta)$ are all based on the condition: the parameter vector $\beta$ is given. Then, the estimation method proposed in the next section depends on a profile method, which needs an iterative estimation method to implement the estimation procedure. In this subsection, we introduce a separable representation for generalized linear form to avoid the theoretical and computational complexity.

For the generalized linear regression functions $f_j(\beta^T(x-\mu))$ in (\ref{generalized-weight-nonlinear}),
we consider a particular choice of $\beta$
as
$\beta=c_h\Sigma^{-1}\sigma_h $, where $h(y):y\in\Omega\mapsto R$ is a given measurable function, and the constant $c_h\neq 0$ depends on $h$. The choices of $h$ and $c_h$ will be given later.
To demonstrate this structural assumption, we provide the following motivating examples.

{\it Example 4}.
Under linear Euclidean regression (\ref{(linear-model)}), the regression function is
$m(x)=E[Y]+ \beta^T(x-\mu)$. It can be seen that the least square representation of $\beta$ is $\beta=\Sigma^{-1}\sigma$ with $\sigma=(\sigma^{(1)},\cdots,\sigma^{(p)})^T$ and $\sigma^{(j)}=E[Y(X^{(j)}-\mu^{(j)})]$. Then the regression function can be rewritten as $m(x)=E[Y]+\sigma^T\Sigma^{-1} (x-\mu)$. In this simple case, $\beta=c_h\Sigma^{-1}\sigma_h$ with $c_h=1$ and $\sigma_h=\sigma$.

{\it Example 5}. Under an Euclidean space, if regression is expressed by standard orthogonal basis $\{\varphi_1(x),\cdots,\varphi_p(x)\}$, i.e.,
$m(x)=E[Y]+\sum_{j=1}^{p}\theta^{(j)}\varphi_j(x)$, then it can be seen that $\theta^{(j)}=\sigma^{(j)}$ and $\mu^{(j)}=0$ if $X^{(j)}$ is replaced by $\varphi_j(X)$. In this case, we also have $\beta=c_h\Sigma^{-1}\sigma_h$ with $c_h=1$ and $\sigma_h=\sigma$.

{\it Example 6}. Consider the following generalized linear Euclidean regression function:
$m^{\cal N}(x)=E[Y]+\sigma^{(j)} g(\beta^T(x-\mu))$ with $\sigma^{(j)}=E[Y(X^{(j)}-\mu^{(j)})]\neq 0$ for a fixed $j$, where $g(\cdot)$ is a given nonlinear link function.
The existing literature (see, e.g., \cite{Li:Duan:1989} and \cite{Wangetal:2012}) shows that if $E[X|\beta^TX]$ is a linear function of $\beta^TX$, then $\beta=c_h\Sigma^{-1}\sigma_h$ for a constant $c_h$ that depends on a function $h$. 

Motivated by these observations, we introduce a separable generalized linear weight function as
\begin{equation}\label{(weight-function)}s^{\cal N}(X,x,c_h)=1+\sum_{j=1}^{p}(X^{(j)}-\mu^{(j)})f_j
\left(c_h\sigma_h^T\Sigma^{-1}(x-\mu)\right) \end{equation} for a nonzero constant $c_h$. Here the separability means that $\beta$ has the separable expression $\beta=c_n\Sigma^{-1}\sigma_h$, and the estimation procedures  for parameter vector $\beta$ and the regression function are separable (for the details see the next section). In this case, the estimators of $\sigma_h^T$ and $\Sigma^{-1}$ can be constructed by the data only from Euclidean space. This alternative method applies to situation where there is some linear
trend in the regression function (Wang, et al., 2012), for example, $f_j
\left(\beta^T(x-\mu)\right)=(\beta^Tx+c)^2$ for a constant $c\neq 0$, which contains a linear term $2c\beta^Tx$, although the regression function is really nonlinear; otherwise it is possible that $c_h=0$, for example, $f_j
\left(\beta^T(x-\mu)\right)=(\beta^Tx)^2$, without any linear information.

Then, for a given function $h$ (i.e. given $c_h$), the separable nonlinear Fr\'echet
regression function is formally defined by
\begin{equation}\label{(s-nonlinear-Frechet)}m_\oplus^{\cal N}(x,c_h)=
\argmin_{\omega\in \Omega}E[s^{\cal N}(X,x,c_h)d^2(Y,\omega)] \end{equation} with the separable weight function $s^{\cal N}(X,x,c_h)$ given in (\ref{(weight-function)}). We denote {\it the separable nonlinear Fr\'echet regression by SNLFR for short}.

\section{Estimations and asymptotic properties}

\subsection{Estimation for NLFR and its asymptotic properties}

Here, we first introduce the estimation method for the NLFR function $m_\oplus^{\cal N}(x,\beta)$ defined in (\ref{(nonlinear-Frechet)}).
Let $(X_i,Y_i),i=1,\cdots,n$, be independent and identically (i.i.d.) observations of $(X,Y)$, and $\widehat \mu$ and $\widehat \Sigma$ be the empirical estimators of $\mu$ and $\Sigma$, respectively. For example, the estimators can be chosen as
$\widehat \mu=\frac1n\sum_{i=1}^nX_i$ and $\widehat \Sigma=\frac1n\sum_{i=1}^n(X_i-\widehat \mu)(X_i-\widehat \mu)^T$. Similar to the notations used in the previous section, the estimated weight function is denoted by $\widehat s^{\cal N}(X,x,\beta)=1+\sum_{j=1}^{p}(X^{(j)}-\widehat \mu^{(j)})f_j
(x,\beta)$ or $\widehat s^{\cal N}(X,x,\beta)=1+\sum_{j=1}^{p}(X^{(j)}-\widehat \mu^{(j)})f_j
(\beta^T (x-\mu))$.
Then, given $\beta$,
the estimator of the NLFR function can be obtained by
\begin{equation}\label{(estimate-general)}\widehat m_\oplus^{\cal N}(x,\beta)=
\argmin_{\omega\in \Omega}\frac1n\sum_{i=1}^n \widehat s^{\cal N}(X_i,x,\beta)d^2(Y_i,\omega). \end{equation}

Similar to the representation in Theorem \ref{Theorem 2.1.},
in a Hilbert space and given $\beta$, the estimator of NLFR in (\ref{(estimate-general)}) can be explicitly expressed as
\begin{equation}\label{(H-represention)}\widehat
m^{\cal N}_\oplus(x,\beta)=\widehat\beta^{(0)}+\sum_{j=1}^{p}\widehat\sigma^{(j)}f_j
(x,\beta) \mbox{ or } \widehat
m^{\cal N}_\oplus(x,\beta)=\widehat\beta^{(0)}+\sum_{j=1}^{p}\widehat\sigma^{(j)}f_j
(\beta^T (x-\widehat\mu)),\end{equation} where $\widehat\beta^{(0)}=\frac1n\sum_{i=1}^nY_i$ and $\widehat\sigma^{(j)}=\frac1n\sum_{i=1}^n Y_i(X_i^{(j)}-\widehat\mu^{(j)})$.
The representation ensures that the estimator $\widehat
m^{\cal N}_\oplus(x,\beta)$ is identified as a nonlinear model as well, and its structure matches up to the theoretical version given in Theorem \ref{Theorem 2.1.}. Also, the estimator can be constructed by a weighted sum of $Y_i$ with the estimated nonlinear weight $\widehat s^{\cal N}(X_i,x,\beta)$, that is,
\begin{equation}\label{(H-represention-1)}\widehat
m^{\cal N}_\oplus(x,\beta)=\frac1n\sum_{i=1}^n\widehat s^{\cal N}(X_i,x,\beta)Y_i,\end{equation}  which also matches up to the theoretical version given in Corollary \ref{Corollary 2.1.}.
The above representations are useful for theoretical analysis and numerical calculation.

In the estimation procedure for the NLFR estimator $\widehat
m^{\cal N}_\oplus(x,\beta)$ in (\ref{(estimate-general)}), the parameter vector $\beta$ cannot be estimated separately from the estimation procedure. Motivated by the profile likelihood in semiparametric model (see, e.g., \cite{Severini:Wong:1992}), a profile estimation procedure is then designed as follows:
\begin{enumerate} \item[(1)] Given $\beta$, an initial estimator of $m_\oplus^{\cal N}(x,\beta)$ is given as the solution to the optimization problem in (\ref{(estimate-general)}).
  \item[(2)] Based on the obtained estimator $\widehat m_\oplus^{\cal N}(x,\beta)$, the estimator of $\beta$ can be attained as the solution to the following optimization problem:
$$\widehat \beta_{\widehat m}=
\argmin_{\beta\in \mathscr B}\frac1n\sum_{i=1}^nd^2(Y_i,\widehat m_\oplus^{\cal N}(X_i,\beta)).  $$
\item[(3)] Final estimator of the regression function is $\widehat m_\oplus^{\cal N}(x,\widehat \beta_{\widehat m})$.
\end{enumerate}
Particularly, under the situation where $Y$ is defined in a Hilbert space, the profile algorithm can be designed as follows.
\begin{enumerate} \item[Step] 1. Construct the estimators $\widehat\mu$, $\widehat\beta^{(0)}$ and $\widehat\sigma^{(j)}$ by the aforementioned methods.
\item[Step] 2. Using the representation $\widehat m_\oplus^{\cal N}(X_i,\beta)$ given in (\ref{(H-represention)}) or (\ref{(H-represention-1)}), construct the estimator of parameter $\beta$ by
\begin{equation}
\label{(estimate-beta)}
\widehat \beta=
\argmin_{\beta\in \mathscr B}\frac1n\sum_{i=1}^nd^2(Y_i,\widehat m_\oplus^{\cal N}(X_i,\beta)). \end{equation}
\item[Step] 3. With the estimator $\widehat \beta$ obtained in Step 2, compute
the estimator of the regression function by the following explicit representation
$\widehat
m^{\cal N}_\oplus(x,\widehat\beta)=\widehat\beta^{(0)}+\sum_{j=1}^{p}\widehat\sigma^{(j)}f_j
(x,\widehat\beta)$ or $\widehat
m^{\cal N}_\oplus(x,\widehat\beta)=\widehat\beta^{(0)}+\sum_{j=1}^{p}\widehat\sigma^{(j)}f_j
(\widehat\beta^T (x-\widehat\mu))$.
\end{enumerate}
Note that the solution of parameter estimator $\widehat \beta$ in (\ref{(estimate-beta)}) does not have an exact explicit representation. Then, the numerical iterative solution is required in Step 2.

For the estimation consistency, we only present the conclusions for the case where the regression functions are of the form of $f_j(x,\beta)$. For the case of $f_j(\beta^T(x-\mu))$, the asymptotic theory is similar; the detail is omitted here.
The regularity conditions are given in the Supplementary Materials. These regularity conditions are mainly related to, for example, the existence and uniqueness of the Fr\'echet regression functions and their estimators, and that the expected loss $E[s^{\cal N}(X,x,\beta)d^2(Y,\omega)]$ is significantly minimized at $m_\oplus^{\cal N}(x,\beta)$. Moreover, the existing literature \citep{Petersen:Muller:2019} has proven that the models in Examples 1-3 satisfy these regularity conditions.

\begin{theorem}\label{Theorem 3.1.}  Under the regularity conditions C1-C4 given in the Supplementary Materials,

(i) if
all the regression functions $f_j(x,\beta)$ are uniformly continuous for $x\in R^p$ and $\beta\in \mathscr B$,
then it holds that
\begin{equation}\label{(estimator-consistency)}d_E(\widehat\beta,\beta_*)=o_p(1)  \mbox{ and } d(\widehat m_\oplus^{\cal N}(x,\widehat\beta), m_\oplus^{\cal N}(x,\beta_*))=o_p(1) \end{equation}
for any fixed $x\in R^p$, where $\beta_*=
\argmin_{\beta\in \mathscr B}E[d^2(Y, m_\oplus^{\cal N}(X,\beta))]$.

(ii) if without the uniform continuity, suppose that
all the regression functions $f_j(x,\beta)$ are continuous for $x\in R^p$ and $\beta\in \mathscr B$, and the boundedness of $\|x\|_E\leq B$, $\mbox{diam}(\mathscr B)\leq B$ and $\|X_i\|_E \leq B$ holds for a constant $B>0$ and uniformly for all $i=1,\cdots,n$, then, the above consistency in (\ref{(estimator-consistency)}) is satisfied as well.
\end{theorem}

Furthermore, the following theorem presents the convergence rate.

\begin{theorem}\label{Theorem 3.2.}  (i) If
all the regression functions $f_j(x,\beta)$ are uniformly continuous for $x\in R^p$ and  $\beta\in \mathscr B$, under
the regularity conditions C1-C6 in the Supplementary Materials, for any fixed $x\in R^p$, it holds that
$$d(\widehat m_\oplus^{\cal N}(x,\widehat\beta), m_\oplus^{\cal N}(x,\beta_*))=O_p(n^{-1/(2(\gamma-1))}), $$ where the constant $\gamma>1$ is defined in the regularity condition C6.

(ii) If
all the regression functions $f_j(x,\beta)$ are uniformly continuous for $x\in R^p$ and  $\beta\in \mathscr B$, under the regularity conditions C1-C8, it holds that
$$\sup_{x\in R^p}d(\widehat m_\oplus^{\cal N}(x,\widehat\beta), m_\oplus^{\cal N}(x,\beta_*))=O_p(n^{-1/(2(\alpha'-1))}) $$ for some constant $\alpha'>\alpha$, where $\alpha>1$ is defined in the regularity condition C8. \end{theorem}

If without the condition of uniform continuity, we also can get the same convergence rate for the case of $\|x\|_E\leq B$ and $\mbox{diam}(\mathscr B)\leq B$ (the details are omitted here). The parameters $\gamma$ and $\alpha$ defined in the regularity conditions depend on the model conditions. In some common cases, such as Examples 1-3, the parameters can be definitely determined as $\gamma=\alpha=2$ \citep{Petersen:Muller:2019}. This implies the convergence rates of parameter estimation. It is not surprising to attain such convergence rates, because the models are actually in an underlying parametric framework.

\subsection{Estimation for SNLFR and its asymptotic properties}

Let $\widehat \sigma_h=\frac1n\sum_{i=1}^nh(Y_i)(X_i-\widehat \mu)$. Given $h$ and $c_h$, the estimator of the weight function can be expressed as
$\widehat s^{\cal N}(X,x,c_h)=1+\sum_{j=1}^{p}(X^{(j)}-\widehat \mu _t^{(j)})f_j
\left(c_h\widehat \sigma_h^T\widehat\Sigma^{-1}(x-\widehat\mu)\right)$. Then, for given $h$ and $c_h$, we obtain the estimator of the SNLFR function $m_\oplus^{\cal N}(x,c_h)$ as
\begin{equation}\label{(estimator)}\widehat m_\oplus^{\cal N}(x,c_h)=
\argmin_{\omega\in \Omega}\frac1n\sum_{i=1}^n\widehat s^{\cal N}(X_i,x,c_h)d^2(Y_i,\omega). \end{equation} The main difference from the estimation procedure of the NLFR proposed in the previous subsection is that here we does not need the profile method to estimate the SNLFR function $\widehat m_\oplus^{\cal N}(x,c_h)$, because $c_h$ is an external parameter and independent of the model, and it only depends on the function $h$. Besides, similar to Theorem \ref{Theorem 2.1.}, we have the following representation: under a Hilbert space with the conditions of Theorem \ref{Theorem 2.1.}, it holds that
\begin{equation}\label{(estimation-representation)} \widehat
m^{\cal N}_\oplus(x,c_h)=\widehat\beta^{(0)}+\sum_{j=1}^{p}\widehat\sigma^{(j)}f_j
\left(c_h\widehat\sigma^T_h\widehat\Sigma^{-1}(x-\widehat\mu)\right)=\frac1n\sum_{i=1}^n\widehat s^{\cal N}(X_i,x,c_h)Y_i.
 \end{equation}
The above representations match up to the theoretical versions given in Theorem \ref{Theorem 2.1.} and Corollary \ref{Corollary 2.1.}.

Theoretically, a special condition given in the following theorems is the linearity condition on $X$: $E[X|\beta^TX]$ is a linear function of $\beta^TX$. This linearity condition is satisfied when $X$ has an elliptical distribution (see, e.g., \cite{Li:1991}, \cite{Cook:1998} and \cite{Cook:Ni:2005}).
With these conditions, the following theorem presents the estimation consistency:

\begin{theorem}\label{Theorem 3.3.}  Suppose that $E[X|\beta^TX]$ is a linear function of $\beta^TX$, and all the link functions $f_j(u)$ are continuous for $u\in R$.

(i) Under the regularity conditions C1'-C3' in the Supplementary Materials, for given $h$ and $c_h$, and any fixed $x\in R^p$, it holds that
$d(\widehat m_\oplus^{\cal N}(x,c_h), m_\oplus^{\cal N}(x,c_h))=o_p(1).$

(ii) Under the regularity conditions C1', C2' and C4' in the Supplementary Materials, for given $h$ and $c_h$, it holds that
$$\sup_{\|x\|_E\leq B}d(\widehat m_\oplus^{\cal N}(x,c_h), m_\oplus^{\cal N}(x,c_h))=o_p(1) $$
for some constant $B>0$. In addition to the above conditions, if all the link functions $f_j(u)$ are uniformly continuous for $u\in R$, then,
$$\sup_{x\in R^p}d(\widehat m_\oplus^{\cal N}(x,c_h), m_\oplus^{\cal N}(x,c_h))=o_p(1). $$
\end{theorem}

Furthermore, the following theorem presents the  convergence rate.

\begin{theorem}\label{Theorem 3.4.}  Suppose that $E[X|\beta^TX]$ is a linear function of $\beta^TX$, and all the link functions $f_j(u)$ are continuous for $u\in R$.

(i)  Under
the regularity conditions C1'-C6' in the Supplementary Materials, for given $h$ and $c_h$, and any fixed $x\in R^p$, it holds that
$$d(\widehat m_\oplus^{\cal N}(x,c_h), m_\oplus^{\cal N}(x,c_h))=O_p(n^{-1/(2(\gamma-1))}), $$ where the constant $\gamma>1$ is defined in the regularity condition C6'.

(ii) Furthermore, under the regularity conditions C1'-C8', it holds that
$$\sup_{\|x\|_E\leq B}d(\widehat m_\oplus^{\cal N}(x,c_h), m_\oplus^{\cal N}(x,c_h))=O_p(n^{-1/(2(\alpha'-1))}) $$ for some constants $\alpha'>\alpha$ and $B>0$, where $\alpha>1$ is defined in the regularity condition C8'. In addition to the above conditions, if all the link functions $f_j(u)$ are uniformly continuous for $u\in R$, then,  $$\sup_{x\in R^p}d(\widehat m_\oplus^{\cal N}(x,c_h), m_\oplus^{\cal N}(x,c_h))=O_p(n^{-1/(2(\alpha'-1))}).  $$
\end{theorem}



Before ending this subsection, we discuss how to choose the function $h(Y):Y\in\Omega\mapsto R$ and how to identify the constant $c_h$. It can be seen from the procedure of estimating the SNLFR function that we need the function $h(Y)$ only for independently identifying the parameter vector $\beta$ in the link functions $f_j(\beta^T(x-\mu))$ with the special choice of $\beta$ as $\beta=c_h\Sigma^{-1}\sigma_h$. That is to say that after identifying $\beta=c_h\Sigma^{-1}\sigma_h$, the estimation procedure for $\beta$ is mostly independent of the metric space data, making the estimation procedure easy to implement.

As shown in Example 6, the constant $c_h$ depends only on the chosen function $h$, implying that we should first choose $h$ and then determine the constant $c_h$. As stated in the existing literature \citep{Wangetal:2012}, a common choice of $h(\cdot)$ is $h(y)=G(y)$, the distribution function of $Y$, if it is well-defined in a metric space. When the distribution $G(y)$ is unknown in practice, we may use the empirical distribution function $\widehat G(y)$ of $Y$. If without the distribution in a metric space, $h(Y)$ can be chosen as a one-to-one correspondence between the metric space $\Omega$ and a subset of $R$. Then, for a given $h$,
an optimal choice of the constant $c_h$ is defined by
\begin{equation}\label{(ch)}c^*_h=
\argmin_{c_h\in R}\frac1n\sum_{i=1}^nd^2(Y_i,\widehat m_\oplus^{\cal N}(X_i,c_h)). \end{equation}
By the representation Theorem \ref{Theorem 2.1.}, the SNLFR function $\widehat m_\oplus^{\cal N}(x,c_h)$ is uniquely identified when $h$ is given. It guarantees that the above optimal choice of $c^*_h$ is unique as well under a Hilbert space.
 The ideal optimality (\ref{(ch)}) leads to the following empirical version:
\begin{equation}\label{(ch-empirical)}c^e_h=
\argmin_{c_h\in \mathscr C}\frac1n\sum_{i=1}^nd^2(Y_i,\widehat m_\oplus^{\cal N}(X_i,c_h))\end{equation} for a given finite subset $\mathscr C\subset R$.

More generally, the optimal strategy of simultaneous choices for both $h(\cdot)$ and $c_h$
will be given in the Supplementary Materials.

\section{Numerical analyses}

The main simulation studies are reported in this section, and some additional numerical results will be given in the Supplementary Materials.
In the simulation experiments, two types of responses are considered, the first one is the one-dimensional probability distributions and the second one is symmetric positive definitive matrices. 
For a comprehensive comparison, we consider the two competitors, the linear Fr\'echet regression (LFR), also known as global Fr\'echet regression, and the local Fr\'echet regression (LoFR) \citep{Petersen:Muller:2019}. The computations for the LFR and the LoFR can be carried out by R package \texttt{frechet} \citep{Chenetal:2020}. Following \cite{Petersen:Muller:2019}, a grid of bandwidths $\{0.05,0.1,\ldots,0.3\}$ together with the Gaussian kernel are used for constructing LoFR, and the optimal bandwidth is selected by 10-fold cross-validation.


For given the independently generated testing data $\{\widetilde{X}_i, \widetilde{Y}_i\}_{i=1}^{\widetilde{n}}$ and a specific metric $d(\cdot,\cdot)$, we adopt two types of mean squared errors
$\mathrm{MSE}_{Y}=\frac{1}{\widetilde{n}}\sum_{i=1}^{\widetilde{n}}
    d^2(\widehat{m}_\oplus(\widetilde{X}_i),\widetilde{Y}_i)$ and
$\mathrm{MSE}_{m}=\frac{1}{\widetilde{n}}\sum_{i=1}^{\widetilde{n}}d^2
(\widehat{m}_\oplus(\widetilde{X}_i),m_\oplus(\widetilde{X}_i))$
 to evaluate the performance of each method,
where $\widehat{m}_\oplus(\cdot)$ is an estimated Fr\'echet regression function based respectively on the SNLFR, the NLFR, the LFR and the LoFR. In the following simulations, all the results are obtained by computing the averaged values of $\mathrm{MSE}_{Y}$ and $\mathrm{MSE}_{m}$ with 100 replications. In addition, the quality of the parameter estimator of $\beta$ in the NLFR is measured by the averaged squared error across replications defined by
$
\mathrm{ASE}_{\beta}=\frac{1}{100}\sum_{r=1}^{100}\|\widehat{\beta}^{[r]}-\beta\|_2^2,
$
where $\widehat{\beta}^{[r]}$ is the parameter estimator in the $r$th replication.

\subsection{Fr\'echet regression for probobility distributions}

Let $\Omega$ be the metric space of probability distributions on $R$ with finite second-order moments, equipped with the Wasserstein metric distance $d_W^2(\cdot,\cdot)$ (see Example 1 for a detailed introduction). In this subsection, the responses $Y$ represent the distribution function with the quantile function $Q(Y)$. Following \cite{Petersen:Muller:2019}, for the sake of notational simplicity, we also denote the quantile function corresponding to $Y$ as $Y$.  To achieve a computationally manageable implementation of the Wasserstein metric distance, the discrete approximation of $d_W^2(G_1,G_2)$ is computed as $m^{-1}\sum_{i=1}^m(G_1^{-1}(t_i)-G_2^{-1}(t_i))^2$ based on an equally spaced grid $\{t_1,\ldots, t_m\}$ on the interval $[0,1]$ with $m=20$ \citep{Tuckeretal:2023}.

The regression function is given by
$
m_\oplus(x)=E(Y(\cdot)|X=x)=U_0+\alpha^\mathrm{T}g(x)+(V_0+\alpha^\mathrm{T}g(x))(\Phi^{-1}(\cdot)+1),
$
where $\Phi(\cdot)$ is the standard normal distribution function, $\alpha=(\alpha_1,\ldots,\alpha_p)^\mathrm{T}\in R^p$, $U_0, V_0\in R$ and $g(x)=(g_1(x, \beta),\ldots,g_p(x, \beta))^\mathrm{T}$ with the known component  functions $g_1,\ldots,g_p$.
For the quantile function, the random response $Y$ is generated by
adding noise as follows: $Y(\cdot)=U+V (\Phi^{-1}(\cdot)+1)$ with $U \mid X \sim N(U_0+\alpha^\mathrm{T}g(X), v_1)$ and $V \mid X \sim \operatorname{Gamma}((V_0+\alpha^\mathrm{T}g(X))^2 / v_2, v_2 /(V_0+\alpha^\mathrm{T}g(X)))$.

We next provide some implementation details in the SNLFR and the NLFR, such as the determination of the function $h(Y)$ and the link function $f(x)=(f_1(x,\beta),\ldots,f_p(x,\beta))^T$.
To better reflect the information of $h(Y)$, we choose $h(Y)=U-E(U)+(V-E(V))(\int_0^{1}\Phi^{-1}(t) dt +1)=U+V-E(U+V)$  for the SNLFR. Note that based on an equally spaced grid $\{t_1,\ldots, t_m\}$ on $[0,1]$, a discrete representation of the response $Y$ is $(Y(t_1),\ldots,Y(t_m))^\mathrm{T}$. Thus, the function $h(Y)$ can be calculated by $Y^{(m)}-E[Y^{(m)}]$  with $Y^{(m)}=m^{-1}\sum_{i=1}^mY(t_i)$. Here, the expectation $E[Y^{(m)}]$ can be approximately computed by its empirical estimator given a random sample. The optimal choice of $c_h$ is further derived via (\ref{(ch-empirical)}) in Section 3.2, where the \texttt{optim} function in the R program can be used with the option ``Brent''.
Moreover, we discuss how to choose the link function $f(x)$. Let $\sigma=(\sigma^{(1)},\ldots,$ $\sigma^{(p)})^\mathrm{T}$. By the representation Theorem \ref{Theorem 2.1.}, it is readily shown that $\beta^{(0)}=U_0+\alpha^ \mathrm{T}E[g(X)]+(V_0+\alpha^ \mathrm{T}E[g(X)])(\Phi^{-1}(\cdot)+1)$ and $\sigma=\mathrm{Cov}(X,g(X))(2\alpha+\alpha\Phi^{-1}(\cdot))$, where $\mathrm{Cov}(X,g(X))=E[(X-\mu)(g(X)-E[g(X)])^\mathrm{T}]$. According to $m_\oplus(x)=\beta^{(0)}+\sigma^ \mathrm{T} f(x)$, the link function $f(x)$ can be determined by
$\mathrm{Cov}(g(X),X)f(x)=g(x)-E[g(X)].$

In the following, we compare different Fr\'echet regression procedures under various models and the settings of sample sizes and dimensions.

\noindent{\textbf{Model 1.1.}}  (\textit{Linear Fr\'echet Regression}) Consider $(X^{\prime (1)},\ldots,X^{\prime(p)})\sim N_p(0,\Sigma)$ with $\Sigma$ $=(0.5^{|i-j|})_{p\times p}$, and then set $X^{(j)}=2\Phi(X^{\prime(j)})-1$ for $j=1,\ldots,p$. Clearly, $X^{(j)}\sim \mathcal{U}(-1,1)$ for each $j$. Let $\alpha=(1,0,\ldots,0)^\mathrm{T}$, $g(x)=(\beta^\mathrm{T}(x-\mu),0,\ldots,0)^\mathrm{T}$ and  $f(x)=(f_1(x,\beta),0,\ldots,0)^\mathrm{T}$. Consider $p\in\{2,5\}$ and $n\in\{100,200,500\}$. For $p=2$, we choose $\beta=(1,-0.5)^\mathrm{T}$ and $V_0=2$. For $p=5$, we choose $\beta=(1,-0.5,2,1.5,-1)^\mathrm{T}$ and $V_0=6.5$,
the additional parameters are set as $U_0=0$, $v_1=1$, $v_2=0.5$ and $\widetilde{n}=500$.

\noindent{\textbf{Model 1.2.}}  (\textit{Generalized linear Fr\'echet Regression}) Let $(X^{(1)},\ldots,X^{(p)})\sim N_p(0,\Sigma)$ with $\Sigma=(0.5^{|i-j|})_{p\times p}$ and $g(x)=((\beta^\mathrm{T}(x-\mu)+1)^2,0,\ldots,0)^\mathrm{T}$. The additional parameter $V_0$ is set as 0.5. The other settings are the same as in Model 1.1.


\noindent{\textbf{Model 1.3.}}  (\textit{Nonlinear Fr\'echet Regression}) Let $g(x)=(g_1(x,\beta),0,\ldots,0)^\mathrm{T}$ with $g_1(x,\beta)=\sum_{j=1}^{p_1}(\beta^{(j)}(x^{(j)}-\mu^{(j)})+1)^2+\sum_{j=p_1+1}^{p}\exp{(\beta^{(j)}(x^{(j)}-\mu^{(j)}))}$, where $\beta^{(j)}$ is the $j$th element in $\beta$. Consider the case of $p_1=1$ for $p=2$ and $p_1=3$ for $p=5$. The other settings are the same as in Model 1.2.

For Models 1.1--1.3 with $p=2$, the averaged values of $\mathrm{MSE}_{Y}$ and $\mathrm{MSE}_{m}$ with standard errors in parentheses over 100 repetitions are reported on Table \ref{table1 p2}. It can be seen from Table \ref{table1 p2} that the SNLFR, the NLFR and the LFR yield similar $\mathrm{MSE}_{Y}$ and $\mathrm{MSE}_{m}$ values for Model 1.1, and moreover, these $\mathrm{MSE}_{Y}$ and $\mathrm{MSE}_{m}$ values are smaller than those of the LoFR, especially when the sample size is small. For  Model 1.2 with a generalized linear structure, the NLFR has the best performance with the SNLFR coming
in a close second. The performance of LFR is the worst for Model 1.2 since the LFR is only suitable for linear F\'rechet regression. For the nonlinear model Model 1.3, the NLFR still maintains the best performance and significantly outperforms the LFR and the LoFR under both the $\mathrm{MSE}_{Y}$ and $\mathrm{MSE}_{m}$ measures.

\begin{table}[htbp]
\caption{\label{table1 p2} The averaged $\mathrm{MSE}_{Y}$ and $\mathrm{MSE}_{m}$ of various methods and the associated standard errors (in
parenthesis) for Models 1.1--1.3 with $p=2$.} \vspace{0cm}
\linespread{0.9}
 \centering
\resizebox{1\textwidth}{!}{
\small{\begin{tabular}{lllccccccccc}
\toprule
& &  & \multicolumn{4}{c}{$\mathrm{MSE}_{Y}$}            & \multicolumn{4}{c}{$\mathrm{MSE}_{m}$}  \\
\cmidrule(r){4-7}   \cmidrule(l){8-11}
& & &SNLFR  &NLFR &LFR  &LoFR  &SNLFR  &NLFR &LFR  &LoFR \\\hline
&\multirow{6}{*}{Model 1.1}
&$n=100$  & \textbf{1.914}   & \textbf{1.914}   & 1.916   & 2.063   & \textbf{0.054}   & \textbf{0.054}   & 0.055   & 0.206   \\
&&      & (0.011) & (0.012) & (0.011) & (0.015) & (0.004) & (0.004) & (0.004) & (0.010) \\
&&$n=200$  & 1.912   & \textbf{1.911}   & 1.912   & 1.981   & \textbf{0.025}   & \textbf{0.025}   & \textbf{0.025}   & 0.096   \\
&&      & (0.011) & (0.010) & (0.010) & (0.011) & (0.002) & (0.002) & (0.002) & (0.003) \\
&&$n=500$  & \textbf{1.896}   & \textbf{1.896}   & \textbf{1.896}   & 1.923   & \textbf{0.011}   & \textbf{0.011}   & \textbf{0.011}   & 0.039   \\
&&      & (0.011) & (0.011) & (0.011) & (0.011) & (0.001) & (0.001) & (0.001) & (0.001) \\ \hline
&\multirow{6}{*}{Model 1.2}
&$n=100$  & 2.127   & \textbf{1.979}   & 7.424   & 5.143   & 0.234   & \textbf{0.085}   & 5.516   & 3.244   \\
&&      & (0.023) & (0.015) & (0.099) & (0.586) & (0.022) & (0.005) & (0.092) & (0.583) \\
&&$n=200$  & 1.982   & \textbf{1.928}   & 7.101   & 3.144   & 0.091   & \textbf{0.039}   & 5.248   & 1.256   \\
&&      & (0.016) & (0.010) & (0.088) & (0.133) & (0.010) & (0.002) & (0.088) & (0.132) \\
&&$n=500$  & 1.918   & \textbf{1.898}   & 7.148   & 2.509   & 0.036   & \textbf{0.015}   & 5.262   & 0.682   \\
&&      & (0.012) & (0.011) & (0.092) & (0.146) & (0.003) & (0.001) & (0.084) & (0.147) \\ \hline
&\multirow{6}{*}{Model 1.3}
&$n=100$  & --   & \textbf{1.984}   & 13.051   & 7.087   & --   & \textbf{0.102}   & 11.222   & 5.227   \\
&&      & -- & (0.012) & (0.172) & (1.753) & -- & (0.006) & (0.172) & (1.763) \\
&&$n=200$ & --   & \textbf{1.917}   & 12.529   & 4.399   & --   & \textbf{0.043}   & 10.549   & 2.513   \\
&&      & -- & (0.011) & (0.171) & (0.943) & -- & (0.002) & (0.161) & (0.936) \\
&&$n=500$  & --   & \textbf{1.882}   & 11.887   & 2.237   & --   & \textbf{0.015}   & 10.012   & 0.367   \\
&&      & -- & (0.010) & (0.165) & (0.038) & -- & (0.001) & (0.163) & (0.037) \\
\bottomrule
\end{tabular}}}
\end{table}

The simulation results in the case of $p=5$ are summarized in Table \ref{table1 p5}. It can be observed that although the $\mathrm{MSE}_{Y}$ and $\mathrm{MSE}_{m}$ values of NLFR are larger relative to the case of $p=2$, they remain the smallest for Model 1.1--1.3 among all competitors. It is noteworthy that LFR shows quite poor performance compared to NLFR for Models 1.2--1.3, indicating that it is not suitable to fit generalized linear and nonlinear structures. What needs to be explained is that SNLFR produces larger $\mathrm{MSE}_{Y}$ and $\mathrm{MSE}_{m}$ values than those of the NLFR for Model 1.2. It is because the  function $h(Y)$ causes a loss of the information of $Y$ on covariates, especially when the sample size is small and the dimensionality of covariates is high. In addition, as the increase of sample size, the $\mathrm{MSE}_{Y}$ and $\mathrm{MSE}_{m}$ values of the SNLFR remarkably decrease, which shows that the SNLFR is also effective in dealing with the generalized linear structures. Considering the curse of dimensionality for the LoFR, the package \texttt{frechet} only handles cases where the dimension of $X$ is not larger than 2. For the case of $p=5$, the simulation results of the LoFR are not shown in Table \ref{table1 p5}.

\begin{table}[htbp]
\caption{\label{table1 p5} The averaged $\mathrm{MSE}_{Y}$ and $\mathrm{MSE}_{m}$ of various methods and the associated standard errors (in
parenthesis) for Models 1.1--1.3 with $p=5$.} \vspace{0cm}
\linespread{0.9}
 \centering
\resizebox{1\textwidth}{!}{
\small{\begin{tabular}{lllccccccccc}
\toprule
& &  & \multicolumn{4}{c}{$\mathrm{MSE}_{Y}$}            & \multicolumn{4}{c}{$\mathrm{MSE}_{m}$}  \\
\cmidrule(r){4-7}   \cmidrule(l){8-11}
& & &SNLFR  &NLFR &LFR  &LoFR  &SNLFR  &NLFR &LFR  &LoFR \\\hline
&\multirow{6}{*}{Model 1.1}
&$n=100$  & 1.996   & \textbf{1.986}   & \textbf{1.986}   & --   & 0.125   & \textbf{0.119}   & \textbf{0.119}   & --   \\
&&      & (0.013) & (0.012) & (0.013) & -- & (0.007) & (0.006) & (0.006) & -- \\
&&$n=200$  & 1.921   & \textbf{1.917}   & \textbf{1.917}   & --   & 0.061   & \textbf{0.056}   & \textbf{0.056}   & --   \\
&&      & (0.010) & (0.010) & (0.010) & -- & (0.003) & (0.002) & (0.002) & -- \\
&&$n=500$  & 1.894   & \textbf{1.892}   & \textbf{1.892}   & --   & 0.024   & \textbf{0.023}   & \textbf{0.023}   & --   \\
&&      & (0.010) & (0.010) & (0.010) & -- & (0.001) & (0.001) & (0.001) & -- \\ \hline
&\multirow{6}{*}{Model 1.2}
&$n=100$  & 288.349   & \textbf{2.248}   & 740.977   & --   & 286.633   & \textbf{0.368}   & 739.117   & --   \\
&&      & (26.216) & (0.086) & (14.123) & -- & (26.212) & (0.083) & (14.182) & -- \\
&&$n=200$  & 141.654   & \textbf{2.047}   & 697.442   & --   & 139.878   & \textbf{0.171}   & 695.946   & --   \\
&&      & (11.447) & (0.045) & (10.926) & -- & (11.471) & (0.044) & (10.883) & -- \\
&&$n=500$  & 56.771   & \textbf{1.917}   & 694.413   & --   & 54.915   & \textbf{0.049}   & 692.475   & --   \\
&&      & (5.012) & (0.012) & (12.014) & -- & (5.005) & (0.003) & (12.002) & -- \\ \hline
&\multirow{6}{*}{Model 1.3}
&$n=100$  & --   & \textbf{2.913}   & 346.129   & --   & --   & \textbf{1.023}   & 344.159   & --   \\
&&      & -- & (0.148) & (29.707) & -- & -- & (0.149) & (29.661) & -- \\
&&$n=200$ & --   & \textbf{2.594}   & 306.084   & --   & --   & \textbf{0.680}   & 304.674   & --   \\
&&      & -- & (0.103) & (5.95) & -- & -- & (0.099) & (5.938) & -- \\
&&$n=500$  & --   & \textbf{2.148}  & 319.947   & --   & --   & \textbf{0.279}   & 318.232  & --   \\
&&      & -- & (0.034) & (8.611) & -- & -- & (0.034) & (8.661) & -- \\
\bottomrule
\end{tabular}}}
\end{table}

\begin{figure}[htbp]
  \begin{center}
   \begin{minipage}{4.6cm}
    \includegraphics[width=1.2\linewidth]{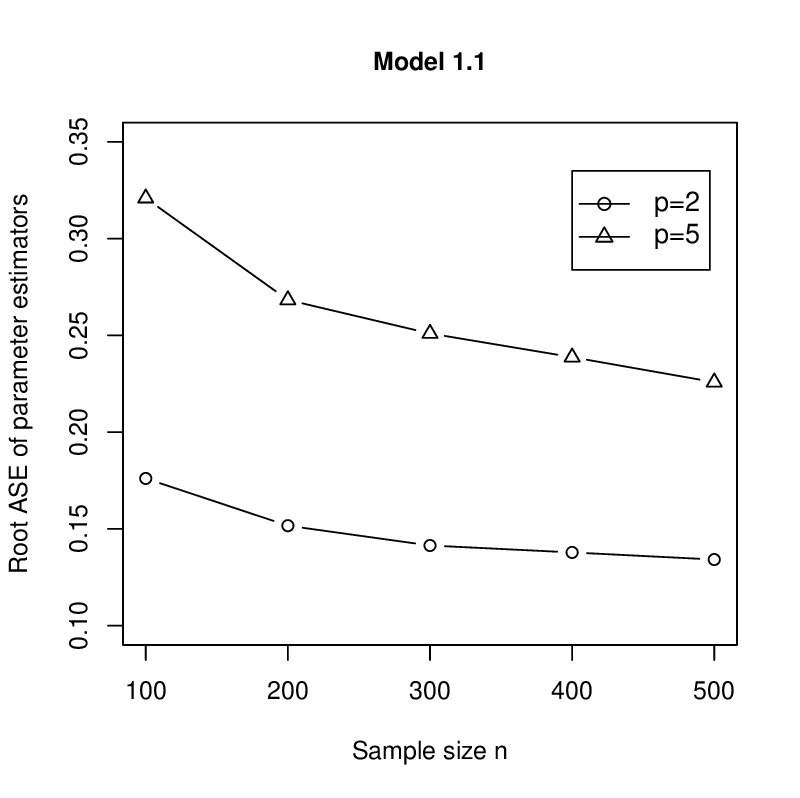}
    \end{minipage}
    \quad
    \begin{minipage}{4.6cm}
    \includegraphics[width=1.2\linewidth]{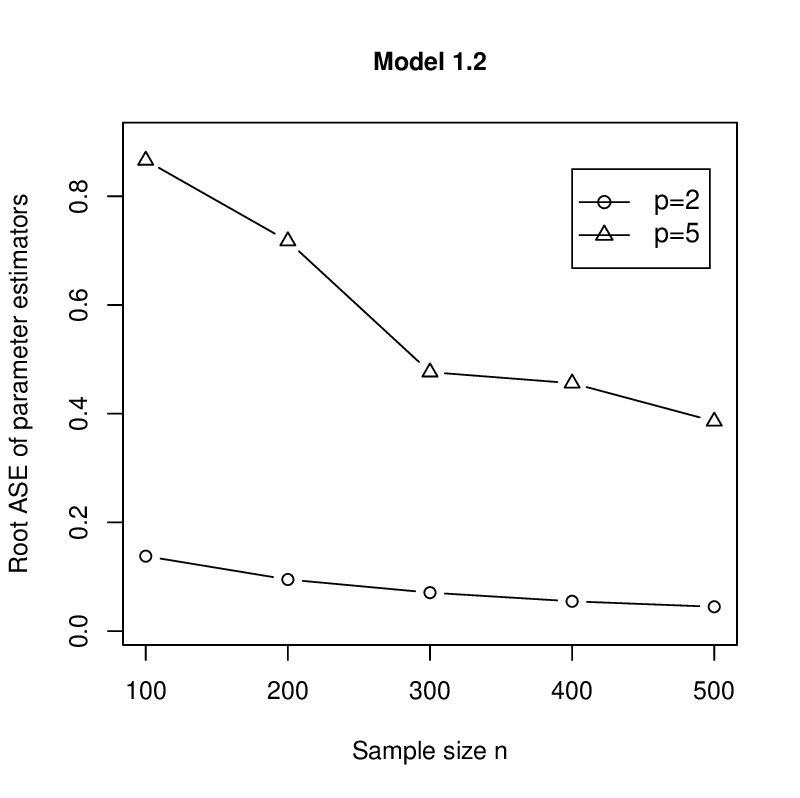}
    \end{minipage}
    \quad
    \begin{minipage}{4.6cm}
    \includegraphics[width=1.2\linewidth]{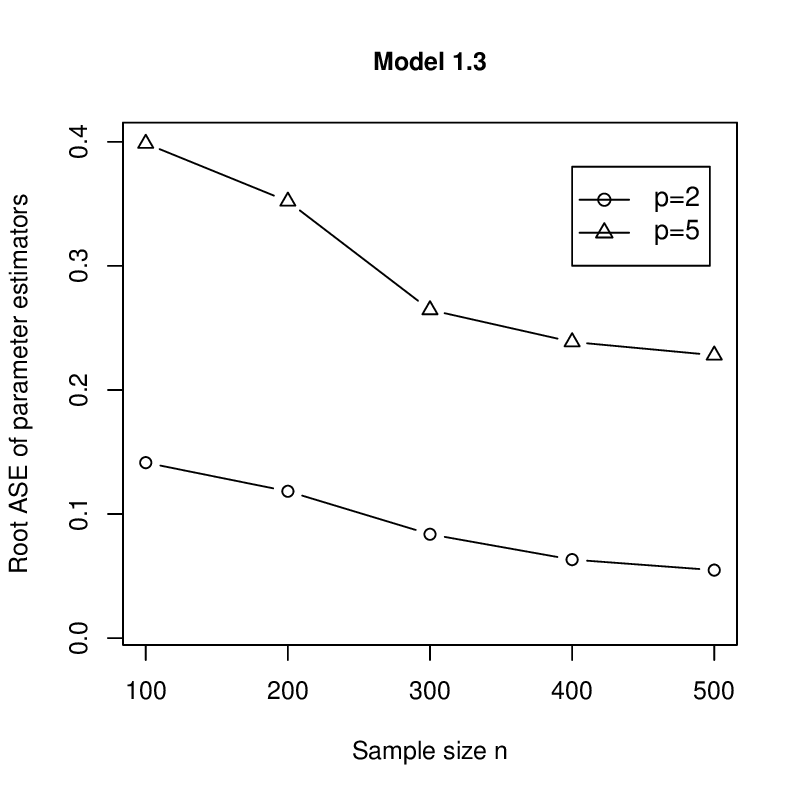}
    \end{minipage}
    \caption{The root $\mathrm{ASE}_{\beta}$ curves of the estimators $\widehat{\beta}$ in NLFR for Models 1.1--1.3.}
  \label{figure beta1}
  \end{center}
  \vspace{-0.5cm}
\end{figure}

Additionally, we report the performance of the estimator $\widehat{\beta}$ in the NLFR for Models 1.1--1.3 in terms of $\mathrm{ASE}_{\beta}$ in Figure \ref{figure beta1}.
 It can be seen that the $\mathrm{ASE}_{\beta}$ values of $\widehat{\beta}$ decrease and approach 0 with the increase of sample size $n$ in both cases of $p=2$ and $p=5$, which confirms the consistency conclusion of Theorem \ref{Theorem 3.1.}.
When $p$ is larger, it is expected that a larger sample size is required for the convergence of $\widehat{\beta}$.

\subsection{Fr\'echet regression for symmetric positive definite matrices}

In this subsection, suppose that $\Omega$ is the set of symmetric positive definite (SPD) matrices of a fixed dimension $m$ equipped with the Frobenius metric $d_F$. Here, in evaluating the performance of various methods, we simultaneously consider the Cholesky decomposition metric $d_C$ for a comprehensive comparison. The Cholesky decomposition metric $d_C$ is defined as follows: For $P_1, P_2\in \Omega$, we can derive $P_1=(P_1^{1/2})^\mathrm{T}P_1^{1/2}$ and $P_2=(P_2^{1/2})^\mathrm{T}P_2^{1/2}$ based on Cholesky decomposition with upper triangle matrices $P_1^{1/2}$ and $P_2^{1/2}$. Then the Cholesky
decomposition distance between $P_1$ and $P_2$ is defined as $d_C(P_1,P_2)=\|P_1^{1/2}-P_2^{1/2}\|_F$, where $\|\cdot\|_F$ is the Frobenius norm.

The regression function is specified by
$
m_\oplus(x)=E(Y|X=x)=(U_0+\alpha^\mathrm{T}g(x))I_m+(V_0+\alpha^\mathrm{T}g(x))P_m,
$
where $U_0+\alpha^\mathrm{T}g(x)>0$, $V_0+\alpha^\mathrm{T}g(x)>0$, $I_m$ denotes the $m\times m$ identity matrix and $P_m$ is a given $m\times m$ positive semi-definite matrices. In our simulation, $P$ can be given for each replication by the following procedure: (1) let $Z$ be a given $m\times m$ matrix with independent $N(0,1)$ elements; (2) compute the orthonormal basis vectors $z_1,\ldots,z_{m^{\prime}}$ of $Z$ with $m^{\prime}=\mathrm{rank}(Z)$; (3) obtain an $m\times m$ matrix $Q$ with $Q=(z_1,\ldots,z_{m^{\prime}})$ for $m^{\prime}=m$ and $Q=(z_1,\ldots,z_{m^{\prime}},0,\ldots)$ for $m^{\prime}<m$; (4) $P=Q^{\mathrm{T}}DQ$ with $D=\mathrm{diag}\{1,\ldots,m\}$.
For symmetric positive definite matrices, the random response $Y$ is generated by adding noise as follows: $Y=U^2I_m+VP_m$ with $U \mid X \sim N((U_0+\alpha^\mathrm{T}g(X)-v_1)^{1/2}, v_1)$ and $V \mid X \sim \operatorname{Gamma}((V_0+\alpha^\mathrm{T}g(X))^2 / v_2, v_2 /(V_0+\alpha^\mathrm{T}g(X)))$. The transformation function $h(Y)$ in the SNLFR is chosen as $h(Y)=\mathrm{trace}(Y-E(Y))$, where
$\mathrm{trace}(\cdot)$ denotes the trace of a matrix. In this way,  we hope that $h(Y)$ retains the information of $U^2$ and $V$ as much as possible.  In simulation, $E(Y)$ is approximately calculated by its empirical estimator based on a given sample. Further, an optimal $c_h$ can be selected based on (2.12). In addition, by the representation theorems, we have $\beta^{(0)}=(U_0+\alpha^\mathrm{T}E[g(X)])I_m+(V_0+\alpha^\mathrm{T}E[g(X)])P_m$ and $\sigma=\mathrm{Cov}(X,g(X))\alpha(I_m+P_m)$. Using $m_\oplus(x)=\beta^{(0)}+\sigma^ \mathrm{T} f(x)$, the link function $f(x)$ is determined by $\mathrm{Cov}(g(X),X)f(X)=g(X)-E[g(X)]$.


We next give Models 2.1-2.3 with the responses are the SPD matrices to assess the performance of each method under different settings.

\noindent{\textbf{Model 2.1.}}  (\textit{Linear Fr\'echet Regression}) Let $(X^{\prime (1)},\ldots,X^{\prime(p)})\sim N_p(0,\Sigma)$ with $\Sigma$ $=(0.5^{|i-j|})_{p\times p}$, and then set $X^{(j)}=2\Phi(X^{\prime(j)})-1$ for $j=1,\ldots,p$. Set $\alpha=(1,0,\ldots,0)^\mathrm{T}$, $g(x)=(\beta^\mathrm{T}(x-\mu),0,\ldots,0)^\mathrm{T}$ and  $f(x)=(f_1(x,\beta),0,\ldots,0)^\mathrm{T}$. Let $p\in\{2,5\}$ and $n\in\{100,200\}$. For $p=2$, $\beta=(1,-0.5)^\mathrm{T}$, $U_0=3$ and $V_0=2$. For $p=5$, $\beta=(1,-0.5,2,1.5,-1)^\mathrm{T}$, $U_0=8$ and $V_0=6.5$.
The additional parameters are set as $m=3$, $v_1=1$, $v_2=0.5$ and $\widetilde{n}=500$.

\noindent{\textbf{Model 2.2.}}  (\textit{Generalized linear Fr\'echet Regression}) Let $(X^{(1)},\ldots,X^{(p)})\sim N_p(0,\Sigma)$ with $\Sigma=(0.5^{|i-j|})_{p\times p}$ and $g(x)=((\beta^\mathrm{T}(x-\mu)+1)^2,0,\ldots,0)^\mathrm{T}$. The additional parameters are set as $U_0=1.5$ and $V_0=0.5$. The other settings are the same as in Model 2.1.


\noindent{\textbf{Model 2.3.}}  (\textit{Nonlinear Fr\'echet Regression}) Let $g(x)=(g_1(x,\beta),0,\ldots,0)^\mathrm{T}$ with $g_1(x,\beta)=\sum_{j=1}^{p_1}(\beta^{(j)}(x^{(j)}-\mu^{(j)})+1)^2+\sum_{j=p_1+1}^{p}\exp{(\beta^{(j)}(x^{(j)}-\mu^{(j)}))}$. Set $p_1=1$ for $p=2$ and $p_1=3$ for $p=5$. The other settings are the same as in Model 2.2.

To save space, the simulation results of various methods for Models 2.1--2.3 are recorded in Tables \ref{table2 p2}--\ref{table2 p5} (with Cholesky decomposition metric) and Figures S1--S2 (with Frobenius metric) in Section S.1 of the Supplementary Materials. For linear fr\'echet regression Model 2.1,  under the Cholesky decomposition and Frobenius metric, the NLFR and SNLFR perform with almost indistinguishable differences and they are both superior to the LFR and the LoFR, especially for the small sample size. Not surprisingly, we can observe from Tables \ref{table2 p2}--\ref{table2 p5} and Figures S1--S2 that the NLFR is the best performer for the generalized linear and nonlinear models, and its superiority is more apparent than other methods when $p$ is larger. The SNLFR has a satisfactory performance when $p$ is smaller and the difference between the NLFR and the SNLFR is insignificant when the sample size is large for $p=5$.

\begin{table}[htbp]
\caption{\label{table2 p2} The averaged $\mathrm{MSE}_{Y}$ and $\mathrm{MSE}_{m}$ of various methods and the associated standard errors (in
parenthesis) for Models 2.1--2.3 with $p=2$ under the Cholesky decomposition metric $d_C$.} \vspace{0cm}
\linespread{0.9}
 \centering
\resizebox{1\textwidth}{!}{
\small{\begin{tabular}{lllccccccccc}
\toprule
& &  & \multicolumn{4}{c}{$\mathrm{MSE}_{Y}$}            & \multicolumn{4}{c}{$\mathrm{MSE}_{m}$}  \\
\cmidrule(r){4-7}   \cmidrule(l){8-11}
& & &SNLFR  &NLFR &LFR  &LoFR  &SNLFR  &NLFR &LFR  &LoFR \\\hline
&\multirow{6}{*}{Model 2.1}
&$n=100$  & \textbf{2.295}   & \textbf{2.295}   & 2.312   & 2.409   & \textbf{0.975}   & \textbf{0.975}   & 1.034   &1.133   \\
&&      & (0.044) & (0.044) & (0.046) & (0.047) & (0.041) & (0.041) & (0.043) & (0.044) \\
&&$n=200$  & \textbf{2.245}   & \textbf{2.245}   & 2.267   & 2.315   & \textbf{0.920}   & 0.921   & 0.986   & 1.037   \\
&&      & (0.050) & (0.050) & (0.052) & (0.052) & (0.046) & (0.046) & (0.048) & (0.048) \\
&&$n=500$  & \textbf{2.198}   & \textbf{2.198}   & 2.223   & 2.238   & \textbf{0.897}   & \textbf{0.897}   & 0.965   & 0.981   \\
&&      & (0.046) & (0.046) & (0.048) & (0.048) & (0.042) & (0.042) & (0.044) & (0.044) \\ \hline
&\multirow{6}{*}{Model 2.2}
&$n=100$  & 2.639   & \textbf{2.612}   & 3.121   & 6.478   & 1.204   & \textbf{1.187}   & 1.731   & 5.079   \\
&&      & (0.060) & (0.060) & (0.063) & (1.083) & (0.058) & (0.058) & (0.059) & (1.081) \\
&&$n=200$  & 2.566   & \textbf{2.558}   & 3.091  & 3.101   & 1.151   & \textbf{1.147}   & 1.715   & 1.736   \\
&&      & (0.049) & (0.048) & (0.049) & (0.079) & (0.047) & (0.047) & (0.048) & (0.079) \\
&&$n=500$  &2.508   & \textbf{2.495}   & 3.001   & 2.822   & 1.101   & \textbf{1.093}   & 1.646   & 1.467   \\
&&      & (0.056) & (0.056) & (0.057) & (0.082) & (0.053) & (0.053) & (0.055) & (0.083) \\ \hline
&\multirow{6}{*}{Model 2.3}
&$n=100$  & --   & \textbf{3.108}   & 4.001   & 4.381   & --   & \textbf{1.847}   & 2.731   & 3.114   \\
&&      & -- & (0.088) & (0.089) & (0.173) & -- & (0.084) & (0.085) & (0.171) \\
&&$n=200$ & --   & \textbf{3.043}   & 3.929   & 4.198   & --    & \textbf{1.764}   & 2.650   & 2.952   \\
&&      & -- & (0.096) & (0.099) & (0.312) & -- & (0.093) & (0.095) & (0.304) \\
&&$n=500$  & --   & \textbf{3.004}   & 3.846   & 3.334   & --   & \textbf{1.746}   & 2.599   & 2.090   \\
&&      & --- & (0.088) & (0.088) & (0.100) & -- & (0.085) & (0.085) & (0.096) \\
\bottomrule
\end{tabular}}}
\vspace{-0.5cm}
\end{table}

\begin{table}[htbp]
\caption{\label{table2 p5} The averaged $\mathrm{MSE}_{Y}$ and $\mathrm{MSE}_{m}$ of various methods and the associated standard errors (in
parenthesis) for Models 2.1--2.3 with $p=5$ under the Cholesky decomposition metric $d_C$.} \vspace{0cm}
\linespread{0.9}
 \centering
\resizebox{1\textwidth}{!}{
\small{\begin{tabular}{lllccccccccc}
\toprule
& &  & \multicolumn{4}{c}{$\mathrm{MSE}_{Y}$}            & \multicolumn{4}{c}{$\mathrm{MSE}_{m}$}  \\
\cmidrule(r){4-7}   \cmidrule(l){8-11}
& & &GLFR  &NLFR &LFR  &LoFR  &GLFR  &NLFR &LFR  &LoFR \\\hline
&\multirow{6}{*}{Model 2.1}
&$n=100$  & \textbf{4.924}   & \textbf{4.924}   & 5.008   & --   & \textbf{3.684}   & \textbf{3.684}   & 3.780   & --   \\
&&      & (0.166) & (0.166) & (0.170) & -- & (0.160) & (0.160) & (0.164) & -- \\
&&$n=200$ & \textbf{4.840}   & 4.849   & 4.913   & --   & \textbf{3.593}   & 3.600   & 3.679   & --   \\
&&      & (0.164) & (0.164) & (0.166) & -- & (0.159) & (0.160) & (0.164) & -- \\
&&$n=500$  & \textbf{4.317}   & \textbf{4.317}   & 4.386   & --   & \textbf{3.110}   & \textbf{3.110}   & 3.194   & --   \\
&&      & (0.165) & (0.165) & (0.169) & -- & (0.161) & (0.161) & (0.165) & -- \\ \hline
&\multirow{6}{*}{Model 2.2}
&$n=100$  & 17.489   & \textbf{8.138}   &28.766   & --   & 15.780   & \textbf{6.806}   & 26.874   & --   \\
&&      & (0.840) & (0.546) & (0.352) & -- & (0.814) & (0.537) & (0.340) & -- \\
&&$n=200$  & 12.006   & \textbf{6.204}   & 27.694   & --   & 10.507   & \textbf{4.893}   & 25.872   & --   \\
&&      & (0.721) & (0.293) & (0.296) & -- & (0.708) & (0.289) & (0.289) & -- \\
&&$n=500$  & 9.150   & \textbf{7.058}   & 27.590   & --   & 7.730   & \textbf{5.733}   & 25.694   & --   \\
&&      & (0.318) & (0.289) & (0.307) & -- & (0.309) & (0.283) & (0.294) & -- \\ \hline
&\multirow{6}{*}{Model 2.3}
&$n=100$  & --   & \textbf{6.267}   & 15.457   & --   & --   & \textbf{4.996}   &13.994   & --   \\
&&      & -- & (0.222) & (0.230) & -- & -- & (0.217) & (0.229) & -- \\
&&$n=200$  & --   & \textbf{6.078}   & 15.811   & --   & --   & \textbf{4.815}   &14.333   & --   \\
&&      & -- & (0.223) & (0.239) & -- & -- & (0.220) & (0.239) & -- \\
&&$n=500$ & --   & \textbf{5.780}   & 15.245   & --   & --   & \textbf{4.510}   &13.776   & --   \\
&&      & -- & (0.219) & (0.234) & -- & -- & (0.215) & (0.227) & -- \\
\bottomrule
\end{tabular}}}
\vspace{-0.5cm}
\end{table}

Furthermore, the performance of the estimator $\widehat{\beta}$ concerning the $\mathrm{ASE}_{\beta}$ measure for $p=2$ and $p=5$ is exhibited in Figure S3 in Section S.2 of the Supplementary Materials. It can be seen that when the sample size changes from 100 to 500, the $\mathrm{ASE}_{\beta}$ values gradually become smaller and closer to 0 for both $p=2$ and $p=5$.

To sum up, these comprehensive numerical results show that the proposed SNLFR and NLFR methods produce results that are significantly superior to commonly competitive methods for the Fr\'echet regression models with generalized linear and nonlinear structures. Also, these results can well verify the theoretical conclusions about consistency given in the previous sections.

\section{Real data analysis}
In this section, we apply the proposed two approaches to the human mortality dataset across countries.  The age-at-death distributions are considered as random object responses of interest, and the objective is to explore the relationship between the age-at-death distributions and country-specific covariates. The data are obtained from United Nations Databases (\url{http://data.un.org/}) and UN World Population Prospects 2019 Databases (\url{https://population.un.org/wpp/Download}) in the form of life tables for $n=162$ countries in the period 2015-2020.  For the
sake of space, the detailed analysis is presented in Section S.2 of the Supplementary Materials. From these detailed analysis we can see that NLFR outperforms remarkably the baseline LFR and LoFR methods in all scenarios. Also, the SNLFR performs slightly inferior to the LoFR in the case of the small sample size of the training set, but it will gradually outperform the LoFR with the increase of sample size of the training set. Overall, the NLFR procedure is more recommendable in practice.

\setcounter{equation}{0}
\section{Conclusions and future works}

As shown in the introduction, the existing Fr\'echet regression methods identify a model only in a linear framework. Even in the existing nonparametric and semiparametric Fr\'echet regressions, the models are handled by locally or approximately linear techniques, and the resulting models are locally or approximately linear in nature.
We in the previous sections proposed two types of Fr\'echet regressions:
generalized linear and nonlinear Fr\'echet regressions.
These frameworks can be utilized to fit the essentially nonlinear models. Particularly, the SNLFR can return to the standard LFR when the underlying model is indeed linear. Structurally, the nonlinear Fr\'echet objective functions are defined by an unconditional expectation, and the related Euclidean variable and the metric space object are separated into different factors. In particular, under a Hilbert space, the NLFR and SNLFR have explicit representations. These favorable structures make the estimation procedures easy for theoretical analysis and numeral computation.
Moreover, the estimation consistency was established systematically in the previous sections. Some
comprehensive simulation studies and real data analysis were given demonstrating that the estimation methods are easy to use, and the performances of the estimations are significantly better than the competitors.

Although in a Hilbert space, the new methods are computationally simple, its calculation burden is relatively heavy because the profile estimation is involved when $Y$ is in a general metric space. Moreover, unlike the LFR, the variable selection in SNLFR cannot be transformed into a ridge regression framework. Thus, it is difficult to implement the variable selection in SNLFR with high-dimensional covariates, and it is also difficult to extend the methods to more general nonlinear models. These are interesting issues
and are worth further study in the future.

\bibliographystyle{cheng}
\bibliography{ref}

\end{document}